\documentclass[usenatbib]{mn2e}

\usepackage[latin1]{inputenc}
\usepackage{graphicx}
\usepackage{threeparttable}
\begin{document}

\title{{\bf Warm-Hot Intergalactic Medium in the Sculptor Supercluster}}
\author[Zappacosta et al.]
{L.~Zappacosta,$^1$\thanks{Send offprint requests to: zappacos@nabhas.ps.uci.edu} R.~Maiolino,$^2$
  F.~Mannucci,$^3$ R.~Gilli,$^2$ P.~Schuecker,$^4$\\
$^1$ Dipartimento di Astronomia e Scienza dello Spazio, Largo
E. Fermi 2, I-50125 Firenze, Italy \\
$^2$ Osservatorio Astrofisico di Arcetri Largo E. Fermi 5, I-50125
Firenze, Italy \\
$^3$ Istituto di Radioastronomia - CNR Largo E.Fermi 5, I-50125 Firenze 
Italy \\
$^4$ Max-Planck-Institut f\"ur extraterrestrische Physik,
Giessenbachstra\ss e, 85748 Garching, Germany \\
}

%\titlerunning{Warm-Hot Intergalactic Medium in the Sculptor Supercluster}
%\date{Received ..., Accepted ...}
%\normalsize

\maketitle

\begin{abstract}
We have analyzed the soft X-ray emission in a wide area of the
Sculptor supercluster by using overlapping ROSAT PSPC pointings.
After subtraction of the point sources we have found evidence for extended,
diffuse soft X-ray emission. We have investigated the nature of such
extended emission through the cross-correlation with the density of
galaxies as inferred from the M\"unster Redshift Survey.  
In particular we have analyzed the correlation as a 
function of the temperature of the X-ray emitting gas.
We have found 
a significant correlation of the galaxy distribution only
with the softest X-ray emission (0.1--0.3~keV) 
and only for gas temperatures kT $< 0.5$ keV.
We have excluded that this soft X-ray diffuse emission, and its correlation
with the galaxy distribution, is significantly contributed by
unresolved AGN, group of galaxies or individual galaxies.
The most likely explanation is that the soft, diffuse X-ray emission
is tracing Warm-Hot Intergalactic Medium, with temperatures below
0.5 keV, associated with the large-scale structures in the Sculptor
supercluster.
\end{abstract}
\begin{keywords}
Large-scale structure of Universe -- X-rays: diffuse background  
\end{keywords}

\section{Introduction} 
Cosmological simulations predict the formation at low redshifts
($\rm{z}\,< 1$) of a diffuse gas phase with temperatures of 
the order of $T \sim 10^{5.5}\div 10^7 \,\rm{K}$ and typical densities
10--30 times the mean baryonic density (although 30\% of this gas can
exceed overdensities greater than 60, and even greater than 100
in the proximity of clusters of galaxies). This gas phase 
should be distributed in large-scale 
filamentary structures connecting virialized structures 
\citep{cen,dave}.
Such Warm-Hot Intergalactic Medium (WHIM)
has been identified as the main
contributor to the missing matter in the baryonic census,
i.e. $\sim 36 \pm 11$ per cent of the baryons
\citep{fukugita_last}  \footnote{In this value are comprised both the
  low redshift Lyman~$\alpha$ forest and the OVI absorbers whose
indipendent contribution to the cosmic baryonic fraction is still
subject to uncertainties due to the possible double counting of the 
absorbers, since both phases can coexist in the same systems.}.
The formation 
of these warm gaseous filaments is due to the infall of baryonic matter
onto the previously formed dark matter cosmic web. The gravitational
potential of the dark matter heats the gas through shocks 
and triggers the formation of galaxies. 
%Such Warm-Hot Intergalactic Medium (WHIM) 
%can be observed in the soft X-rays (up to $\sim 2 \,\rm{keV}$)
%as low surface brightness structures. 
%The detection of its radiation is hinded because of many Galactic
%foregrounds, such as the Local Hot Bubble (LHB) and the
%Galactic halo, and extragalactic background contributions due to
%groups of galaxies, clusters and AGNs. 
The WHIM
can be observed in the soft X-rays \citep[below $\sim2\,\rm{keV}$;][]{croft}
as low surface brightness structures. 
The detection of its radiation is very difficult because of many Galactic
foregrounds (such as the Local Hot Bubble --LHB-- and the
Galactic halo) and extragalactic background due AGNs, 
groups of galaxies and clusters. Simulations and X-ray
background studies have shown that the WHIM 
continuum emissivity below $2\,\rm{keV}$ is
roughly of the same order of magnitude as the Galactic foregrounds.
More specifically
$\rm F_{0.5-2keV}(WHIM)\approx 7\, keV\,
  s^{-1}\,cm^{-2}\,sr^{-1}\,keV^{-1}$
 \citep[][]{croft,kuntz} and
 $\rm F_{0.2-0.3keV}(WHIM)\approx 15\, keV\,
  s^{-1}\,cm^{-2}\,sr^{-1}\,keV^{-1}$
 (Croft, private communication). Within this context,
\citet{pierre} showed, from simulated observations that {\em XMM} can
observe strong filaments up to $\rm{z}\sim 0.5$ in the 
$0.4-4\,\rm{keV}$ energy band. \\
\citet{cen1995} pointed out that this gas phase should also emit
characteristic spectral lines mainly due to Oxygen, Neon and Iron
ions. The level of emissivity of these spectral features is below the 
sensitivity and spectral resolution limits of the current X-ray instruments.
However, cosmological simulations show that these lines will be detectable
with the future generation of X-ray satellites
\citep{yoshikawa,fang}. \\
Various detections of (continuum) WHIM emission
have been claimed, either obtained by observing
soft X-ray structures in galaxy overdense regions 
\citep{scharf,bagchi,zappacosta}, or by detecting a soft X-ray
excess in clusters of galaxies \citep[][]{kaastra,finoguenov}, 
or in their proximity \citep{tittley,soltan}, or through shadowing effects \citep{bregman}.
These observations have been possible by means of X-ray satellites
very sensitive to low energies ($<$ 1--2~keV), such as {\it ROSAT} and
XMM.\\
Theoretical works had predicted that the WHIM should be detectable
through UV and X-ray absorption lines imprinted on the spectra
of background QSOs \citep{hellsten,perna}.
The detectability of such absorption features does not depend on the
brightness of the filaments but on their  
column density and on the brightness of the background QSO. 
So far several detections have been reported through X-ray and far-UV
absorption lines probing the hot and cool phase of the
WHIM \citep{nicastro,tripp,mathur}.

\begin{figure}
  \begin{center}
    \includegraphics[angle=270, width=0.45\textwidth]{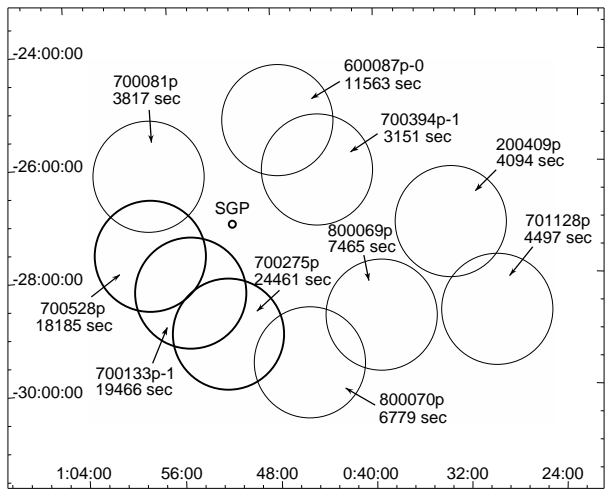}
	\caption{The position of the 10 partially overlapping {\it ROSAT} PSPC
	pointings in the region of the SSC. For each pointing the
	{\it ROSAT} observation ID and the exposure time are shown. 
	The position of the South Galactic Pole (SGP) is also
	shown. The three deepest pointings (that will be used for the 
	subsequent quantitative analysis) are identified with thick circles.}
    \label{exposures}
  \end{center}
\end{figure}

Simulations show that the WHIM should be distributed in
filamentary structures extending over several tens of Mpc and 
connecting clusters of galaxies. Therefore, superclusters (hosting several
clusters) are optimal regions where WHIM is more likely to be detected.
In this work we focus on the Sculptor supercluster
\citep[hereafter SSC,][]{schuecker2,seitter}. 
This is one of the richest local superclusters,
comprising more than 20 clusters  of
galaxies \citep{einasto} spread over a projected length of more than
140 Mpc at a redshift $z \sim 0.105$\footnote{Here and in the rest of
  the paper we will assume a cosmology with $\Omega_{m} = 0.3$, $\Omega_{\Lambda} =
  0.7$ and $\rm{H_{0}} = 70\,Km\,s^{-1}\,Mpc^{-1}$}. It is located in the south
Galactic pole, a region where the Galactic hydrogen column density 
is low enough ($N_H \sim 1.5\pm 0.2\,\times10^{20} \,\rm{cm^{-2}}$ )
to avoid significant effects of patchy absorption  
that could mimic a pattern of apparent X-ray structures 
\citep[see][ for more details]{zappacosta}. 
The SSC has already been observed by \citet{spiekermann} and
\citet{obayashi}, using {\it ROSAT} and ASCA data, 
with the purpose of detecting large-scale X-ray diffuse emission.
They did not find indications for emission extended in large-scale structures.
However, \citeauthor{spiekermann} focused the analysis to the relatively
energetic bands at $0.5-3$~keV (without investigating the correlation with
the galaxy distribution), whereas 
\citeauthor{obayashi} observed in an even harder band ($0.8-10$~keV) and
by using pointings centered on clusters with the
goal of detecting hot diffuse emission in their outskirts.

In this paper we present evidence for a correlation
between the galaxy distribution and soft X-ray emission in the central 
region of the SSC. In particular we show that galaxies and 
the softest X-ray emission ($< 0.3$~keV) correlate in regions with 
gas temperatures kT$< 0.5$~keV. 
This finding is interpreted as WHIM emission associated with the 
large-scale structures in the SSC.

\section{Data Description} %X-ray from {\it ROSAT}, M\"unster Redshift Survey
Our aim is to detect the WHIM over the central region of the
supercluster 
($8.3\times6.4\;\rm{deg^2}$ corresponding to $57\times44 \rm{\:Mpc 
\:at \:z=0.105}$)
which is populated by more than 15 Abell clusters. We considered 
10 PSPC partially overlapping pointings taken from the {\it ROSAT} archive
and which cover the core of the SSC (see Fig.~\ref{exposures}). 
Unfortunately, the exposures of these pointings are not homogeneous,
ranging from 
$\sim\!4$ to $\sim\!24 \,\rm{\: ksec}$. As a consequence, our 
 quantitative analysis will be restricted only to the three deepest
fields (i.e. those with exposures $> 18 \,\rm{ksec}$, represented with
thick circles in Fig.~\ref{exposures}).

To map the galaxy density distribution in the SSC we have used data
from the M\"unster Redshift Project 
\citep[MRSP,][]{seitter,muenster,ungruhe2}. 
The MRSP is a catalog of galaxies obtained by scanning direct
and very low-dispersion objective-prism Schmidt plates over a wide
region ($5000 \,\rm{deg^2}$) in the south Galactic hemisphere.
We have produced a catalog of all objects 
identified as galaxies to a limiting magnitude $\rm{r_{F}} \sim
20.5$\footnote{Galactic extinction in this region
  affects r$_F$ by at most 0.04 magnitudes.}.
The MRSP catalog is suitable to investigate the SSC just because the
photographic plates, where redshifts have been estimated
\citep{ungruhe}, efficiently sample objects at redshift $z=0.1$
\citep[see the redshift histograms in][]{muenster,ungruhe}. 

\begin{figure*}
  \begin{center}
    \includegraphics[angle=270, width=0.98\textwidth]{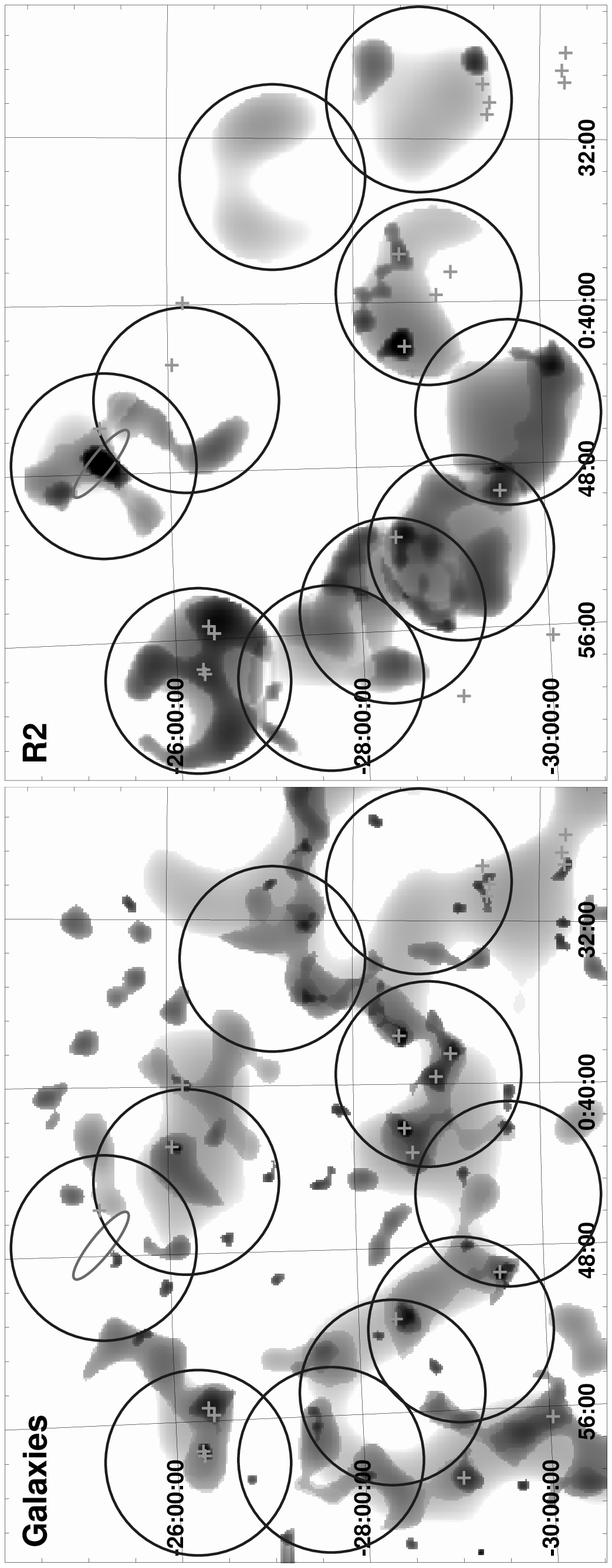}
    \includegraphics[angle=0, width=0.32\textwidth]{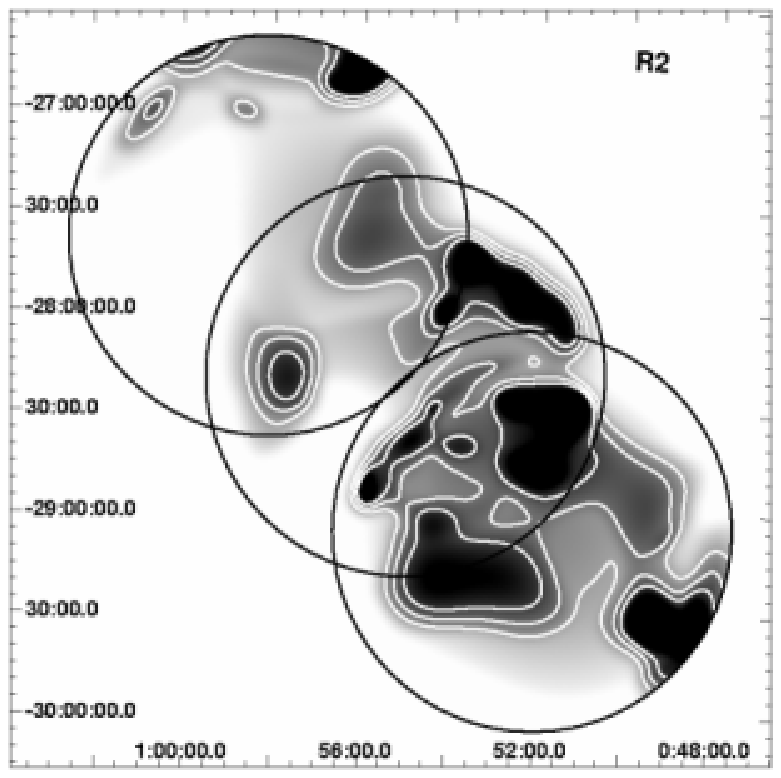}
    \includegraphics[angle=0, width=0.32\textwidth]{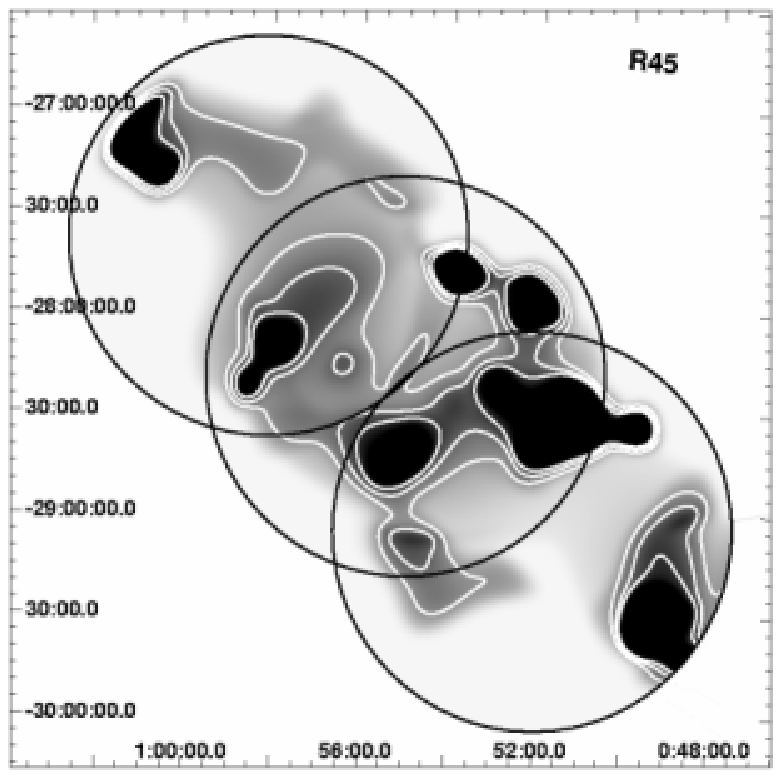}
    \includegraphics[angle=0, width=0.32\textwidth]{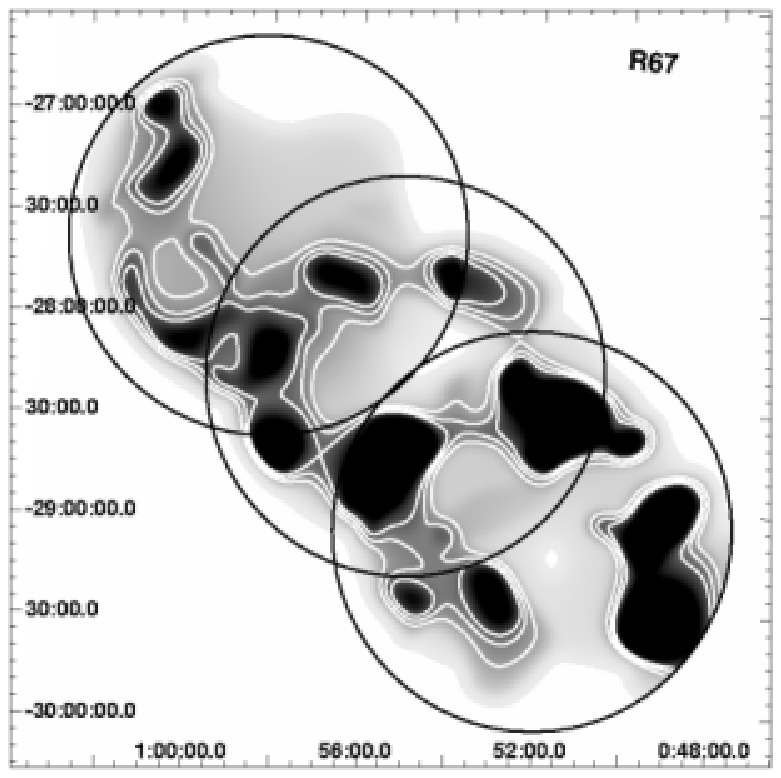}
\caption{
Comparison between structures detected in the distribution of 
galaxies and in the X-ray {\it ROSAT} bands. Structures are found
with a wavelet algorithm and are significant at $4\sigma$ and their extension
followed down to $1 \sigma$.
The top panels show the projected density of galaxies (left) and R2 flux (right)
over the whole field covered by the 10 {\it ROSAT}
pointings. 
Crosses represent
clusters members of SSC and the ellipse at
the top indicates the location of the large starburst galaxy NGC 253.
The bottom panels show (from left to right) the R2, R45 and R67 band
maps for the three deepest fields; contours indicate significance levels
of $2\sigma$, $3\sigma$ and $4\sigma$.
}
    \label{wavelet}
  \end{center}
\end{figure*}			

X-ray data have been reduced by using the software described in
\citet{snowden}, 
%1994ApJ...424..714S
which allows a careful removal of the instrumental background and 
recovers the effective exposure of each region in the field of
view. This careful treatment of the data is optimized to 
detect faint diffuse structures. Moreover, the high sensitivity 
below 0.3~keV combined with the large field
of view ($\sim2^\circ$~diameter) make the {\it ROSAT} PSPC detector the best
X-ray instrument to detect diffuse soft structures on large scales 
(even superior to XMM).
We have detected and subtracted point sources (and the few known
clusters) from the X-ray image both by using the SExtractor
software \citep{bertin} and by means of a wavelet algorithm
\citep{vikhlinin}. 
The first method identifies point sources in all energy bands by 
increasing the size of
the detection filter as a function of the off-axis angle, to account
for the increasing {\it ROSAT} PSF with off-axis angle. 
The second method makes use of the
wavelet transform to identify structures on various angular 
sizes. The latter procedure does not account for PSF
variations, which were then considered {\it a posteriori}.
The details of both procedures are described in \citet{zappacosta}.
In the rest of the paper we will use the second method for qualitative
considerations and the first one for quantitative analysis.

In order to compare X-ray data with the galaxy catalog we have
generated a map of the projected density of galaxies. 
We have used pixel sizes of $2^\prime \times 2^\prime$ in order to 
have an average number of galaxies per pixel of about $\sim 1$.
Moreover, to obtain a better comparison, we have smoothed the galaxy
density images
with a gaussian kernel that radially increases like the PSPC PSF.

\begin{figure*}
  \begin{center}
    \includegraphics[angle=0, width=0.45\textwidth]{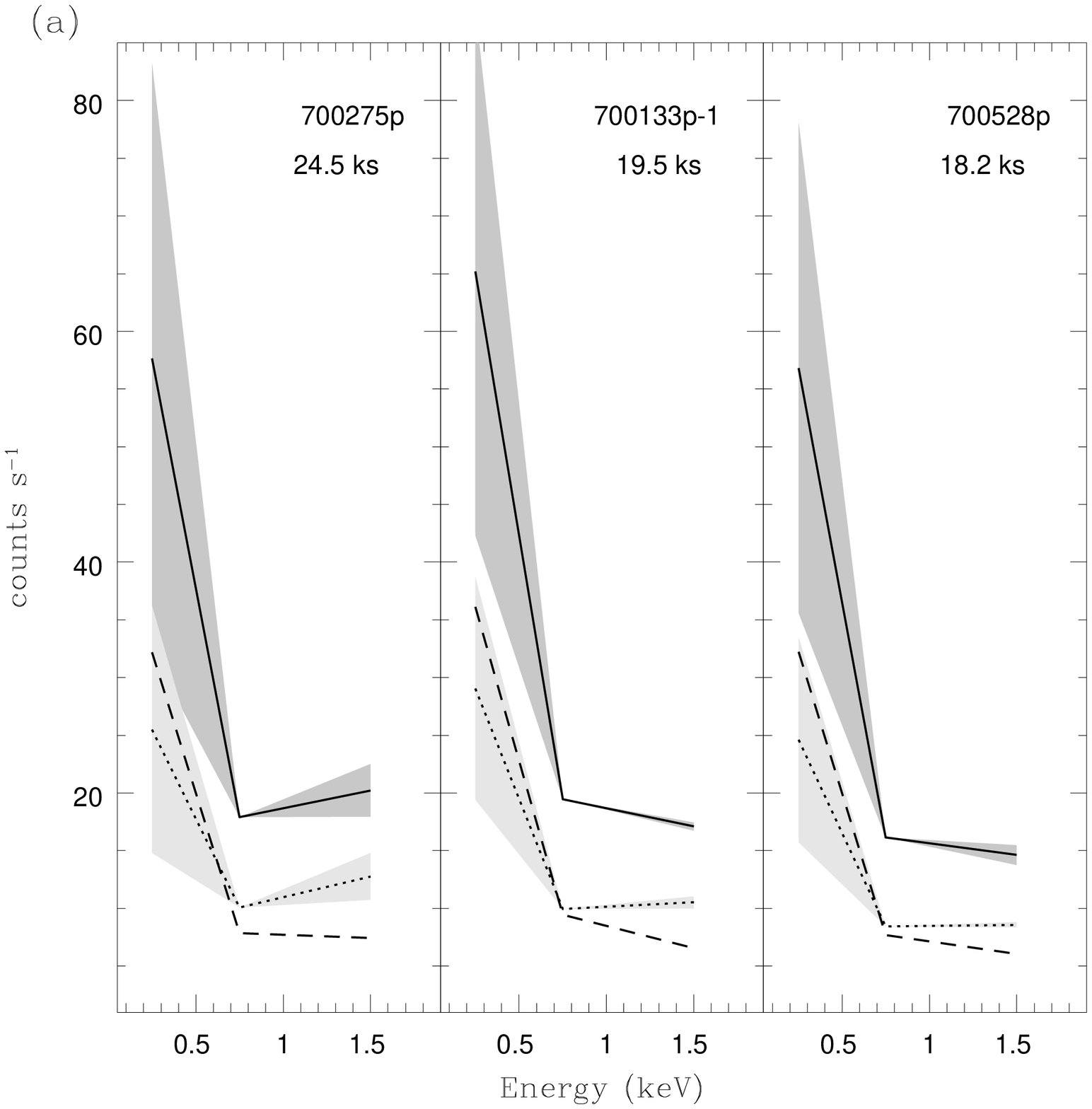}
    \includegraphics[angle=0, width=0.45\textwidth]{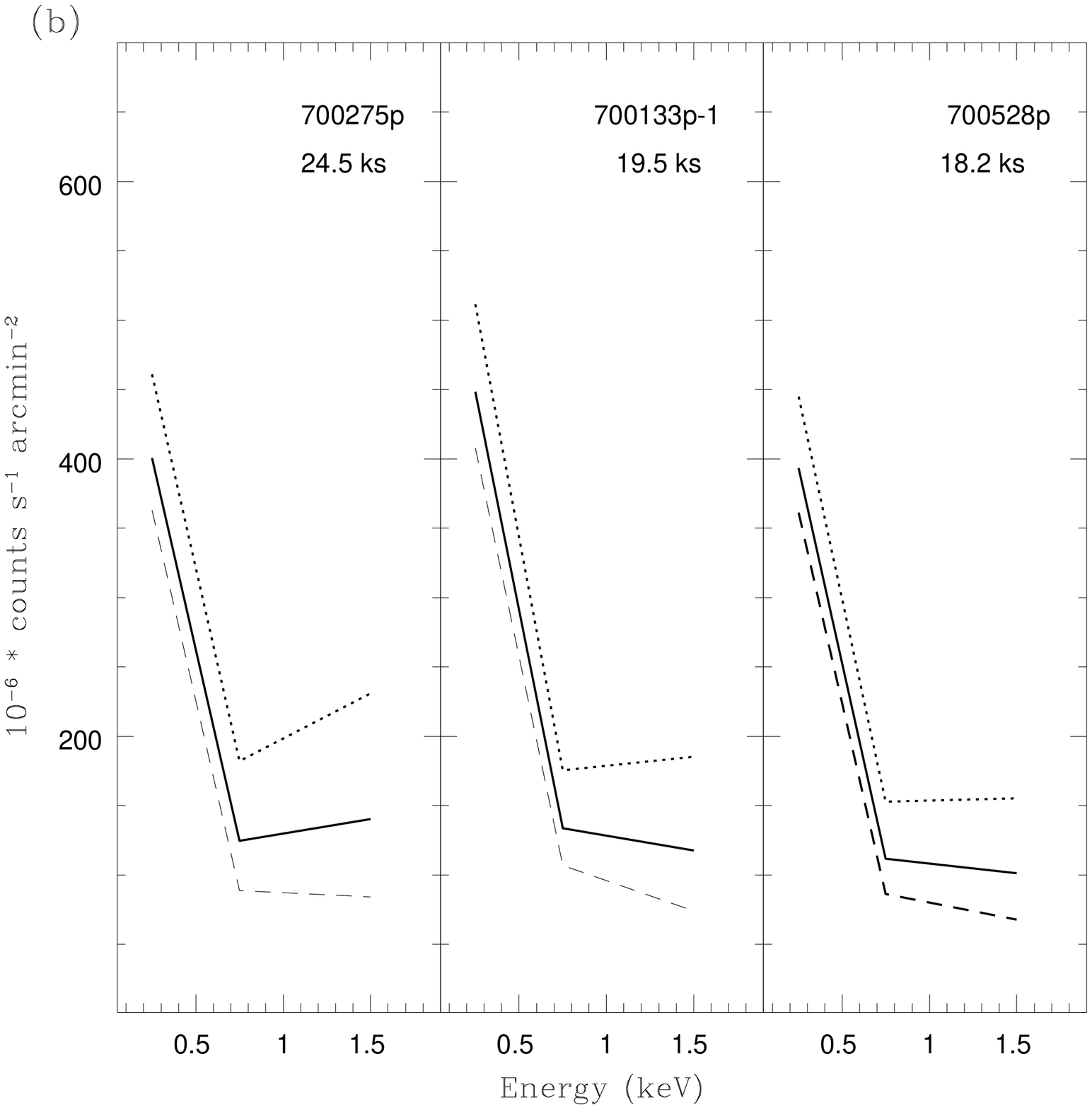}
	\caption{Global spectral shapes (not corrected for Galactic 
	  absorption) for the three 
	deepest fields. Panel a): countrate. Panel b): countrate
	  normalized per unit
	area. The solid lines represent the whole flux detected, the dotted
	  lines show the point sources component and the dashed lines show the
	  residual flux obtained by the subtraction of the first
	  two. Grey regions in panel a) show the dispersions of the spectral
	  shapes over each field.}
    \label{specshape}
  \end{center}
\end{figure*}

\section{A glance at the Sculptor region}
Fig. \ref{wavelet} shows a comparison of the structures found
with the wavelet algorithm both in optical and in the three 
{\it ROSAT} bands and, more specifically (see also Table~\ref{energybands}): 0.14--0.284 keV (R2; the $\frac{1}{4}$~keV band), 0.44--1.21 keV
(R45; the $\frac{3}{4}$~keV band), 0.73--2.04 keV (R67; the 1.5~keV
band).
The upper panels show the projected density of galaxies and the
$\frac{1}{4}$~keV flux for all the pointings. The lower panels show 
the X-ray flux in the three ROSAT bands focused onto the region of the three
deepest
PSPC pointings (see Fig. \ref{exposures}). Contours
at significance levels from
$2\sigma$ to $4\sigma$ (spaced by $1\sigma$) are shown in the latter.

\begin{table}
  \begin{center}
\caption{{\it ROSAT} energy bands and conventional notations.}
   \begin{threeparttable}[b]
    \begin{tabular}{lcc}
\hline
Band & Range & Alt. Notation \\
\hline
R2      & 0.14--0.284 keV & $\frac{1}{4}$~keV\tnote{a}\\
R45     & 0.44--1.21 keV & $\frac{3}{4}$~keV\\
R67     & 0.73--2.04 keV & 1.5~keV\\
\hline
    \end{tabular}
   \begin{tablenotes}
    \item [a] The notation $\frac{1}{4}$~keV is often referred to
    the R12 {\em ROSAT} broad energy band. In the following we will
    use this notation restricted to the R2 energy band.
   \end{tablenotes}
  \end{threeparttable}
\label{energybands} 
 \end{center}
\end{table}
%All the structures have a statistical significance of at least $4
%\sigma$ and their extensions are followed to $1 \sigma$. 
Both galaxies and gas show large-scale structures including 
clusters of galaxies and filamentary structures connecting them.
There is also a lot of X-ray diffuse emission not clearly related to 
visible structures in the galaxy map, which could be due to
foreground by our Galaxy (e.g. LHB and Galactic
halo) or to a blend of unresolved emission by AGNs (mainly in low 
exposures pointings). A contribution could also come
from foreground (or background) superstructures.
Note that \citet{spiekermann} and \citet{obayashi} focused their
searches on the harder X-ray emission (i.e. $> 0.5$~keV) in the region 
of the {\it ROSAT} pointing 800069p (see Fig. 
\ref{exposures}), which is considered the core of the SSC (it contains  
5 Abell clusters).
This region shows
diffuse emission only for the soft $\frac{1}{4}$~keV band
(i.e. at energies below 0.3 keV).

\section{Spectral analysis}\label{spectral_analysis}
Filamentary X-ray structures connecting clusters are not necessarily due to 
warm-hot gas. They could also arise from AGN unresolved emission, as 
both WHIM and galaxies trace gravitational potential wells of dark matter
filaments \citep{scharf,zappacosta}.\\
In order to assess the true nature of these filamentary patterns we
need to perform an accurate comparison with the distribution
of galaxies along with the analysis of the spectral shape of the
X-ray emission. 
We have to limit our analysis to the three deepest fields (700275p,
700133p-1 and 700528p) where we are confident that a larger fraction 
of the AGN contribution to the X-ray background has been resolved.\\
Fig. \ref{specshape} shows the spectral shapes measured for
these pointings sorted by decreasing exposure times. 
Fig. \ref{specshape}{a} shows the countrate, while 
in Fig. \ref{specshape}{b} the spectra are normalized by the area
in each field.
Fluxes are not corrected for the Galactic $\rm{N_H}$ absorption
(this correction would increase the flux in the
$\frac{1}{4}\,\rm{keV}$ band by a factor of $\sim 3$).
Dotted lines show the sum of point-like sources detected 
by SExtractor, while the
solid lines show the total flux in each field (shaded 
regions indicate the dispersions of the slopes within each field).
The difference
between them is the diffuse residual emission and it is shown with the dashed
line. 
We note that the residual soft, diffuse fluxes have values of 
$\sim 380 \times 10^{-6} \rm{counts\,s^{-1}\,arcmin^{-2}}$, well above
those found for the LHB emission through shadowing measurements in the
region of the south Galactic pole \citep{snowden2}, which lie in the range 
$100\,\div\,300 \times 10^{-6} \rm{counts\,s^{-1}\,arcmin^{-2}}$. 
After subtraction of the LHB and after 
correction for Galactic absorption,
the residual diffuse, extended emission has a value of
$ \sim 540 \times
10^{-6} \pm 300 \times 10^{-6} counts\,s^{-1}\,arcmin^{-2}$.
This residual emission may include both Galactic Halo and extragalactic
emission due to either unresolved AGNs, clusters/groups of galaxies and
true WHIM emission. We will identify and disentangle these various
contributions both through an analysis of the spectral shape
and through the correlation with the galaxy distribution.

As discussed above, a residual component due to unresolved
AGNs, or unidentified clusters/groups,
could still be present even in images with longer exposures.
The three {\it ROSAT} energy bands can be used 
to make a color-color diagram, with the goal of separating colors 
typical of a warm gas from unresolved AGNs and clusters. In particular,
the WHIM is expected to have a softer emission with respect to
clusters and AGNs.

\begin{figure}
\begin{center}
  \includegraphics[angle=0, width=0.5\textwidth]{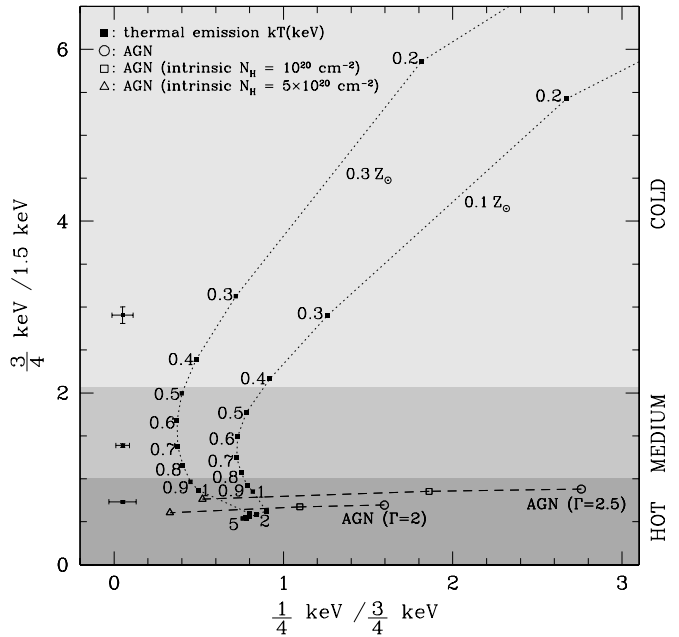}
\caption{Color-color diagram ([$\frac{1}{4}$~keV/$\frac{3}{4}$~keV]
  band ratio versus [$\frac{3}{4}$~keV/1.5~keV] band ratio) for various
  classes of sources. The dotted lines and full symbols indicate
  thermal emission (MEKAL model), corrected for Galactic absorption,
  at redshift $\rm{z}=0.1$ and metallicities
  $\rm 0.1-0.3 Z_{\odot}$, for
  several temperatures expressed in keV. 
  The dashed lines and hollow symbols indicate colors of AGNs with two
  different slopes (quite typical in the ROSAT band), both unabsorbed and
  with intrinsic absorptions of $\rm N_H = 1-5~ 10^{20}~cm^{-2}$.
  The three horizontal shaded regions correspond to the three temperature
  bins adopted in Fig.~\ref{r_col}
  (for each of these regions the error on the average colors
  is shown).
  Note that plasmas with low temperatures (kT$<$0.5~keV) can be efficiently
  selected through high values of the [$\frac{3}{4}$~keV/1.5~keV]
  color (larger than $\sim$2).}
\label{colors}
\end{center}
\end{figure}

Fig. \ref{colors} shows the [$\frac{1}{4}$~keV]/[$\frac{3}{4}$~keV]
band ratio (R2/R45) versus the [$\frac{3}{4}$~keV]/[1.5~keV] band
ratio (R45/R67) for different kind of sources (the ratios
are in counts and corrected for Galactic absorption).
The dotted lines and full symbols indicate the behavior of
thermal emission (optically thin plasma,
MEKAL model),
at a redshift $\rm{z}=0.1$ and 
metallicities of $0.1 \,\rm{Z_\odot}$ and $0.3 \,\rm{Z_\odot}$. 
The dashed lines and hollow symbols indicate colors of AGNs with two
different slopes (quite typical in the ROSAT band), unabsorbed and
with intrinsic absorptions of $\rm N_H = 1-5 \times 10^{20}~cm^{-2}$.
The most interesting feature inferred from the color-color diagram in
Fig.~4 is that the R45/R67 ratio ([$\frac{3}{4}$~keV]/[1.5~keV])
results to be an excellent tracer of the gas temperature.
In particular, gas cooler than $\sim$0.5~keV (typical of
WHIM) is characterized by R45/R67$>$2, while gas warmer than 0.5~keV
(typical of groups and clusters) or AGNs have R45/R67$<$2.
Therefore, we have used the ratio R45/R67
to discriminate WHIM regions from 
areas dominated by unresolved AGNs and clusters.

The X-ray emission in the three fields of the SSC investigated
by us spans a wide range of colors. In particular, R45/R67 ranges from $\sim$0.5
(typical of AGNs and clusters) to values larger than 2 (typical of WHIM).

\section{Correlation Analysis}

The X-ray colors alone (and in particular the inferred low temperatures)
do not necessarily allow the identification of the diffuse X-ray emission
with WHIM, since foreground components like LHB and Galactic Halo
are also characterized by soft emission and low temperatures.
However, any correlation between soft X-ray emission and
distribution of galaxies would support
the idea that the diffuse X-ray emission is extragalactic and
associated with large-scale structures. In this section we discuss
the correlation between X-ray emission and density of galaxies as
a function of the X-ray color.

The most widely used criteria to correlate two data sets 
are the {\em Pearson's correlation coefficient} $\mathbf{r}$ and the 
{\em Spearman's rank correlation coefficient} $\mathbf{r_{s}}$ (also
known as {\em Spearman's rho}). Both assume values ranging from +1 
(perfectly correlated) to -1 (completely anti-correlated). 
A null value means that the two
quantities are not related at all. 
The first correlation coefficient is based
on the assumption that the data follow a gaussian distribution, while the
second makes no assumptions, measuring the correlation on ranked data
(i.e. the data are converted to ranks and then correlated). 
 The Spearman's rho is a 
better indicator that two variables are correlated when they are tied
by a non-linear monotonous correlation.\\
In our case, cosmological models do not predict how galaxies and gas
are linked. More specifically, it is not clear whether there is
a linear correlation between density of galaxies and gas emission, or
some physical mechanism links the formation of galaxies
and the gas phase in a non-linear way. 
Therefore, the Spearman's rho is probably better suited in this case.
However, we will also show the Pearson's correlation
results for comparison.\\
We do not expect to find a high correlation between
galaxies and WHIM. In fact, the three {\it ROSAT} bands do not
correlate strongly among themselves, with correlation
coefficients
in the range $0.2\div0.3$ (that means a low correlation).

Regions containing clusters
would certainly give a higher value of $r_{s}$ because
they have lots of galaxies and high hard X-ray fluxes
in small areas.
Cooler regions should have few galaxies spread over large areas 
with low soft X-ray flux and therefore a low correlation signal
is expected.

We have correlated the galaxy density map with the X-ray merged
maps. Unfortunately, in this analysis
we had to reject the 700528p field since it covers a
region centered at the cross of 4 photographic plates of the
M\"{u}nster Survey, where the
galaxy catalog shows clear spatial inhomogeneities.
So we are left with the two deepest pointings with exposures greater
than 19 ksec (fields 700133p-1 and 700275p).
Moreover,
since we use the ratio R45/R67 to discriminate between ``cold'' and
``hot'' regions, we selected those regions with a good signal in R45.
In particular,
we avoided exceedingly noisy regions by selecting the areas with the
R45 flux higher than $M_X - 2\sigma_X$, where $M_X$ is the median value 
and $\sigma_X$ is the standard deviation over the field (anyhow
this criterion selects most of the field, and more specifically
$\sim95$~per cent).

\begin{figure*}
\begin{center}
\includegraphics[angle=0, width=0.60\textwidth]{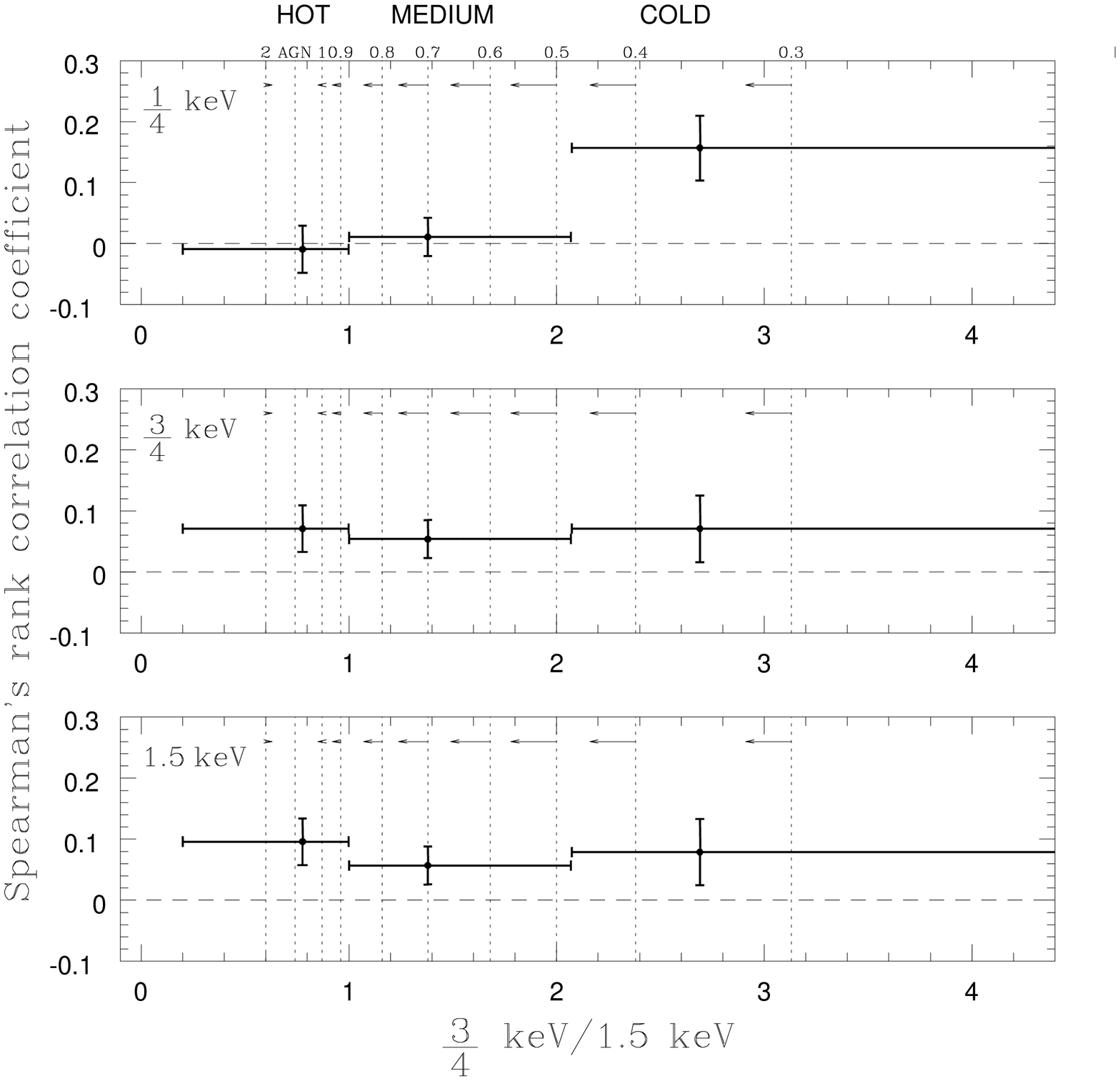}
\includegraphics[angle=0, width=0.60\textwidth]{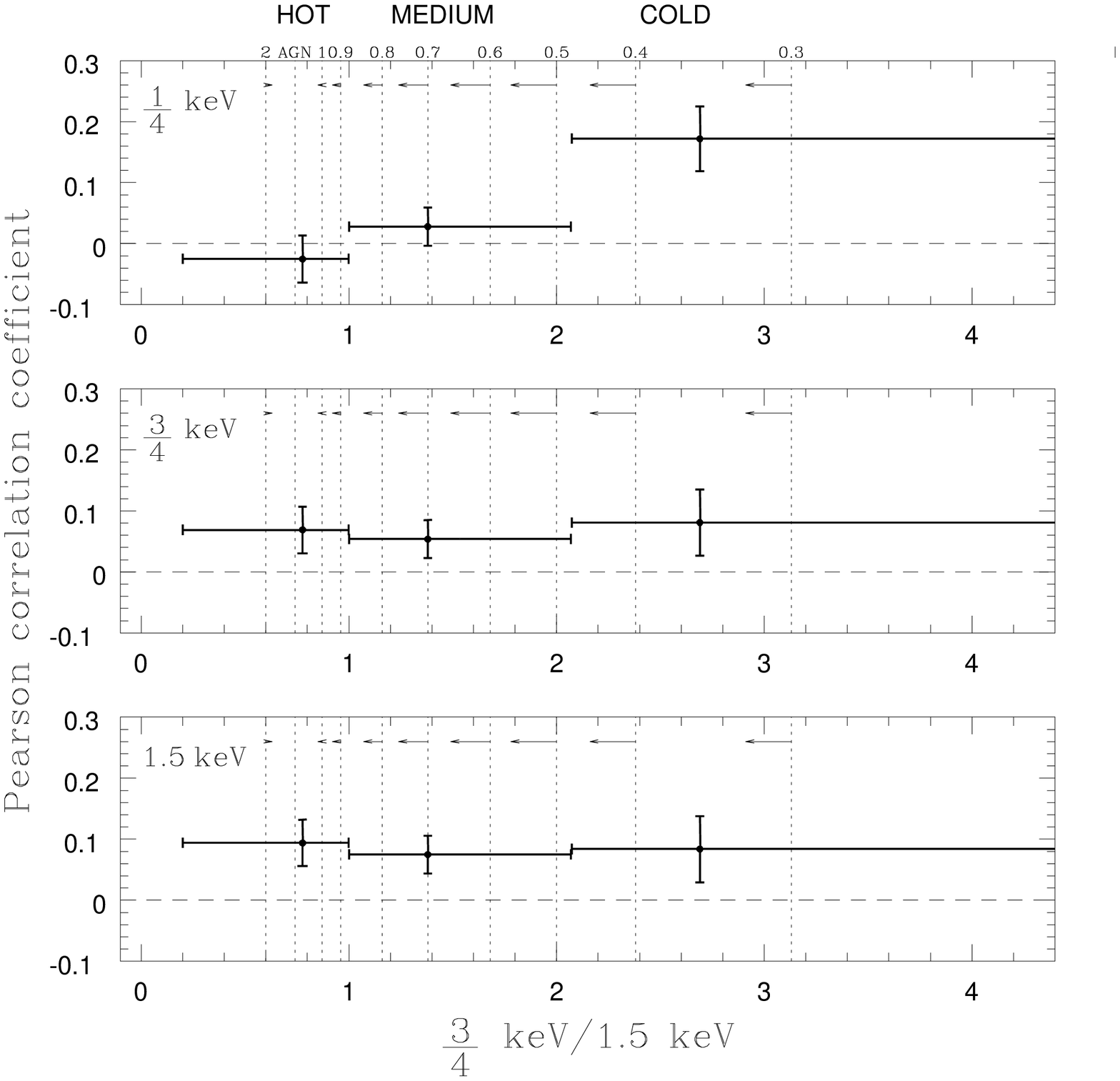}
\caption{Correlation coefficient between galaxy distribution and X-ray
emission, for the three ROSAT bands ($\frac{1}{4}$~keV,
$\frac{3}{4}$~keV, 1.5~keV, from top to bottom), as a function of the
[$\frac{3}{4}$~keV/1.5~keV] ratio.  The vertical dotted lines
indicate the plasma temperatures, as labelled an top of each panel for
a metallicity $0.3\,Z_\odot$, corresponding to specific values of the
[$\frac{3}{4}$~keV/1.5~keV] ratio. Arrows show the shift of the dotted
lines in case of a metallicity of $0.1\,Z_\odot$.  The data are
grouped in three main bins of temperature (``hot'', ``medium'' and
``cold''). For each bin we indicate median value of the
[$\frac{3}{4}$~keV/1.5~keV] ratio. The upper panel is for the
Spearman's correlation coefficient while the lower panel is for the
Pearson's correlation coefficient.}
\label{r_col}
\end{center}
\end{figure*}

Fig. \ref{r_col} shows the behavior of the correlation coefficient
(Spearman's rho and Pearson's coefficients in upper and
lower panels, respectively) 
as a function of the color [$\frac{3}{4}$~keV/1.5~keV] (R45/R67)
for the three {\it ROSAT} energy
bands. The temperatures corresponding to this ratio 
 (assuming a metellicity of $0.3\,\rm{Z_\odot}$)
 are shown in the upper part of the graph 
and indicated in each plot by vertical dotted lines. The arrows show 
the position of the dotted lines in case of a metallicity of
$0.1\,\rm{Z_\odot}$.
We have measured the correlation for three temperature ranges to probe the
``hot'' (kT$>0.9 \,\rm{keV}$), ``medium'' ($0.5-0.9 \,\rm{keV}$)
and ``cold''(kT$<0.5 \,\rm{keV}$) gas. 
For each range we show the median value of the R45/R67 ratio.
Vertical error bars show the $1 \sigma$ confidence on the value of the
correlation coefficient.

The results from the two correlation coefficients do not differ
significantly. 
The most important result is the finding of a weak, but significant
correlation between the density of galaxy and the very soft
$\frac{1}{4}\,\rm{keV}$ flux, {\it only}
for regions with gas temperature below 0.5~keV (i.e. ``cold'' regions).
The correlation coefficient is 0.16--0.17 and significant at
$\rm 3\sigma - 3.3\sigma$ (depending on the correlation coefficient adopted).
The correlations with the other ROSAT bands (and other temperatures)
do not show any significant signal, except for a marginal (2.5$\sigma$)
correlation in the ``hot'' bin of the 1.5~keV band (which will be shortly discussed
at the end of this section).

A correlation between galaxy distribution and very soft X-ray
emission, limited to the
``cold'' (kT$<$0.5~keV) regions, is just what is expected from
WHIM emission which is tracing large-scale structures.
However, there are a few other possibilities that could in
principle explain the correlation in the soft band,
as discussed in the following.

One alternative possibility is that the soft, cold emission
is directly emitted by the individual galaxies
of the SSC, or by a sub-population of them which are particularly active
(starbursts). In this case the X-ray emission would be due to the
warm-hot interstellar medium of the galaxies and to their superwinds.
In our field there are at most $0.5 \,\rm{galaxy\,arcmin^{-2}}$.
We have made two very conservative assumptions: 1) {\it all} these
galaxies are starburst (and not dominated by a mixture of --less
active-- spirals and ellipticals); 2) the residual diffuse X-ray
emission (due to Galactic Halo and extragalactic components) has the
minimum value of $300 \times 10^{-6} \rm{counts\,s^{-1}\,arcmin^{-2}}$
obtained by subtracting to the measured X-ray flux the {\it maximum}
value of the LHB (see Section~\ref{spectral_analysis}). 
We have assumed for a starburst a typical luminosity of 
$10^{41}\,\rm{erg\,s^{-1}}$ in the 0.5--2~keV energy band 
\citep[see Figures 3-4 in][]{norman} and a spectrum made of a
thermal component ($\rm{kT}=0.7$ and $Z=Z_{\odot}$) plus an absorbed
 power-law ($N_H=10^{22}\,\rm{cm^{-2}}$ and $\Gamma = 0.8$)
 representing the X-ray binary contribution \citep[see][]{norman}.
We have estimated that the galaxies should contribute
a flux of $\sim 10^{39}\,\rm{erg\,s^{-1}\,arcmin^{-2}}$ in the R2 ROSAT band.
The latter value is a factor of $\sim 20$ lower than the 
average diffuse R2 flux of $\sim 10^{40.3}\,\rm{erg\,s^{-1}\,arcmin^{-2}}$
that we measure in the ROSAT maps. 
This means that the X-ray emission from normal and starburst
galaxies cannot contribute significantly to the correlation in the
``cold'' regions of the supercluster.

Another possibility, is that cold 
groups of galaxies could contribute in some way to the correlation
in the soft, cold regions. Indeed, a fraction of small groups may
have temperatures as low as 0.4~keV \citep{mulchaey}.
However, as discussed in detail in the Appendix,
the contribution to the coldest gas temperatures due to 
cold groups (i.e. those with kT~$ < 0.5$~keV) is at most the 1 per
cent of the studied region. Therefore, ``cold'' groups cannot
account for the soft X-ray emission nor for the correlation
found in Fig.~4. Warmer systems, such as ``hot'' groups (kT~$ > 0.5$~keV)
and clusters cannot explain the correlation; indeed these systems would give
a significant correlation also in the higher temperature bins and
also in the harder bands.

Finally, another possibility is that the correlation in the soft band,
and at cold temperatures, is contributed by clusters/groups in formation
and not yet virialized. These systems would have a temperature lower
than standard clusters/groups. However, the distinction between forming,
non virialized clusters/groups and WHIM is subtle, and probably just semantic.
Indeed, the definition of WHIM (from a physical point of view) is that
of a medium associated with forming, non-virialized structures \citep{cen}.
Therefore, even a contribution from cold, forming clusters/groups should
be included in the WHIM budget.

Summarizing, the scenario that better explains 
the correlation between galaxies and cold gas emitting at
$\frac{1}{4}\,\rm{keV}$ is that a fraction of the diffuse
soft X-ray emission is due to WHIM
associated with the galaxy distribution in the SSC.

In order to further investigate the latter scenario
we have also tried to estimate the density of the emitting
gas. Such estimate is very uncertain, since we do not have much information
on the geometry of the emitting gas (the WHIM emission is disentangled
only through a statistical analysis over a wide field). Moreover, we do not
know exactly what fraction of the $\frac{1}{4}\,\rm{keV}$ emission is
actually emitted by the WHIM: indeed, although we measure a flux for the
$\frac{1}{4}\,\rm{keV}$ diffuse emission, the weak correlation with the galaxy
distribution may either indicate a real, physically weak association between
galaxies and WHIM, but may also point at a significant dilution from other
unrelated X-ray components (foreground and background emission, Sect.~1).
We have estimated the density of the gas emitting the soft X-ray radiation
by making extreme, opposite assumptions on geometry of the gas
(i.e. either distributed only in the putative filaments of Fig.2, or over the whole field
where the cross-correlation was performed)
and on its contribution to the $\frac{1}{4}\,\rm{keV}$
emission (i.e. either contributing to the whole LHB-subtracted R2 emission, or only
to the 20\% responsible for the correlation with galaxies).
The inferred gas densities range from $\rm 4 \times 10^{-6}~cm^{-3}$ ($\rm \delta \sim 15$),
well in the range of the WHIM specifications, up to
$\rm ~10^{-4}~cm^{-3}$ ($\rm \delta \sim 400$) which may be expected for WHIM
in the proximity of clusters (Sect.~1).

Finally, we shortly discuss the nature of the marginal correlation between
galaxy distribution and X-ray emission in the hard, R67 band (and some of R45),
limited to ``hot'' temperatures (Fig.~5). This can be easily explained in terms
of contribution from a population of unresolved, weakly obscured AGNs.
Indeed, a small absorbing column
density of $\rm{N_{H}} \sim 0.5-1 \times10^{21} \rm{cm^{-2}}$
is enough to absorb most of the $\frac{1}{4}\,\rm{keV}$ flux, while leaving
nearly unaffected the 
harder bands. Moreover, AGNs have R45/R67 colors nearly identical
to ``hot'' plasma (see Fig.~4).
Unresolved, obscured AGNs certainly contribute significantly
to the X-ray background and span a wide range of absorbing $N_{H}$
\citep{mainieri}. Small amounts of X-ray absorption is also detected
in several type 1 AGNs \citep{maiolino,maiolino2} which are the
dominant population found by {\it ROSAT} \citep{lehmann} and {\em
Chandra} \citep{barger,szokoly}. Therefore, it is expected that
a fraction (10-20 per cent) of unresolved AGNs which contribute
to the diffuse signal
detected by us are also slightly absorbed. These slightly obscured,
unresolved AGNs (probably also belonging to the SSC)
are probably responsible for the correlation in the
hard band, and not in R2, for ``hot'' X-ray colors.

%%%%%%%%%

\section{\bf Conclusions} 
We have investigated the emission from Warm-Hot Intergalactic Medium
(WHIM) associated with large-scale structures in the central
region of the Sculptor supercluster ($\rm{z}\,\approx 0.1$).
Ten overlapping {\it ROSAT} PSPC fields, covering the central 8.3$\times$6.4~deg$^2$
of the supercluster, were analysed. After removal of the point sources, the
{\it ROSAT} maps show indication of diffuse, filamentary structures, in some
cases connecting known clusters of the Sculptor. The diffuse emission
spans a wide range of X-ray spectral shapes: from relatively
hard emission expected for clusters and unresolved AGNs,
to very soft emission expected for WHIM.

To investigate the nature of the diffuse X-ray emission we have cross
correlated the X-ray flux with the density of galaxies obtained
from the M\"{u}nster redshift catalog (whose galaxies mostly belong
to the Sculptor supercluster in this region). The correlation has been
analyzed as a function of the gas temperature (or X-ray spectral shape).
The most important result is the finding of
a significant correlation between the diffuse
soft (0.1--0.3~keV) X-ray flux and the density of galaxies at
the coolest gas temperatures (kT$<$0.5~keV). Such a correlation
is interpreted as emission by WHIM associated with the galaxy
distribution.

We have also investigated the possible contribution
to the diffuse soft X-ray emission, and to the correlation with galaxies,
due to individual galaxies and by cold clusters.
We have found that in both cases the contribution is negligible.

We have also detected a weak, marginal correlation between the harder
X-ray flux (1.5~keV, R67 band) and the density of galaxies at apparently higher
gas temperatures (kT$\sim$1~keV). The latter correlation is ascribed to
slightly obscured, unresolved AGNs.

\section*{Acknowledgements}
We thank the referee, F. Nicastro for his comments and helpful suggestions.
We are also grateful to A. Ferrara for very useful comments.
This work was partially supported by the Italian Ministry of Research
(MIUR) and by the Italian Institute of Astrophysics (INAF).

\appendix

\section{The contribution from cold groups of galaxies}

Groups of galaxies are poorly studied objects because of their elusive nature
both in the X-rays and optical. One of the largest samples studied
so far \citep{mulchaey} contains 109 low-redshift galaxy groups. 
It is a collection of
several catalogs of groups selected both in X-ray and
optical. In this catalog temperatures have been derived only for a 
subsample of 61 objects that show also extended diffuse emission. 
The temperatures derived for these clumps range from 1.5~keV down to
0.4~keV. 
The latter overlaps the range of WHIM temperatures. As a consequence,
``cold'' groups could mimic the WHIM behavior, both in terms of
X-ray spectral shape and correlation with galaxies. Therefore, it is important
to quantify the density of ``cold'' clusters and to estimate
their contribution to the soft X-ray emission. The coldest
groups ($\rm{kT} < 0.5 \,\rm{keV}$) 
are also the smallest ($\rm{R_{X}}< 120 \,\rm{kpc}$) and
the least luminous ($\rm{L_{X}} < 10^{41.2} \,\rm{erg/s}$).
To calculate the fraction of area expected to be covered by groups 
we have used the catalog of groups detected by the ESO Slice Project
\citep[ESP;][]{ramella2} in a field close to the SSC region 
($\sim 10 \,\rm{deg}$ away).
The survey has been done in two strips. We
consider only the area near to the SSC (strip A; $22\times1 \,\rm{deg}$). In this 
region they found 190 
groups up to redshift $z = 0.2$, $\sim 71$ at redshifts below
the $\rm{z_{SSC}}$, and only 18 below $z = 0.05$. 
We can assume the extreme scenario
that all groups are colder than 0.5 keV and that all of them
have a size  $\rm{R_{X}} \sim 120 \,
\rm{kpc}$ (which is the maximum size found among cold groups).
With these assumptions, we can calculate what is the maximum fraction
of our field 
occupied by  cold groups in the following redshift ranges: 0--0.05, 
0.05--$\rm{z_{SSC}}$, $\rm{z_{SSC}}$--0.2. To make things simpler we
calculate the angular sizes of groups in these bins using the mean
redshift value,
except for the last one where we assume that all groups have redshift 
$z_{SSC}$.
We exclude from the sample all the identified groups with 3
galaxy members because almost all these groups do not show 
diffuse emission that can contribute to the correlations 
\citep{mulchaey}. With these extreme, conservative assumptions
we obtain that at most 1 per cent of the pixels in our image
could be significantly contaminated
by cold groups emission. This means that cold groups cannot 
contribute to the correlation found in regions with $kT < 0.5 \,\rm{keV}$.

\bibliography{zappacosta}

\end{document}